\documentclass[%
 preprint,
 amsmath,amssymb,
 aps, physrev,
]{revtex4-2}

\usepackage{graphicx}
\usepackage{dcolumn}
\usepackage{bm}

\usepackage[english]{babel} 
\begin{document}
\preprint{APS/123-QED}
\title{\textbf{IMF slope derived from a pure probabilistic model} }
\author{Yu-Fu Shen}
\email{Contact author: shenyf@cho.ac.cn}
\affiliation{Changchun Observatory, National Astronomical Observatories, Chinese Academy of Sciences, Changchun, 130117, China.}%
\date{\today}
\begin{abstract}
The stellar initial mass function is of great significance for the study of star formation and galactic structure. Observations indicate that the IMF follows a power-law form. This work derived that when the expected number of stars formed from a spherical molecular cloud is much greater than 1, there is a relationship between the slope $\alpha$ of the IMF and $r^n$ in the radius-density relation of spherically symmetric gas clouds, given by $\alpha = 3/(n+3)$ ($\Gamma_{\mathrm {IMF}} = n/(n+3)$). This conclusion is close to the results of numerical simulations and observations, but it is derived from a pure probabilistic model, which may have underlying reasons worth pondering.
\end{abstract}
\maketitle

\section{Introduction}

The stellar initial mass function (IMF) is a fundamental concept in astrophysics that quantifies the distribution of masses for newly formed stars. It is crucial for understanding star formation processes, the evolution of galaxies, and the dynamics of stellar populations. The IMF describes how many stars of different masses are created during star formation, typically represented by a power-law function established by Edwin E. Salpeter\cite{1955ApJ...121..161S}, whose work has since been expanded upon by various observations \cite[e.g.][]{annurev:/content/journals/10.1146/annurev-astro-082708-101642,annurev:/content/journals/10.1146/annurev-astro-081811-125528,annurev:/content/journals/10.1146/annurev-astro-032620-020217}. Variations in the IMF across different environments, such as metal-rich versus metal-poor regions, suggest that the conditions under which stars form can significantly alter the mass distribution \cite[e.g.][]{10.1093/mnras/stz103,10.1093/mnras/stac1549,10.1093/mnras/stac2060,10.1093/mnras/stad3581}. Thus, the IMF has profound implications for galactic chemistry and dynamics. Despite the widespread acceptance of the IMF concept, there are many uncertainties, such as stellar evolution \cite[e.g.][]{heger2023encyclopedia}, because the observed stellar mass is not equal to the zero-age main sequence mass. Practical biases and error sources must also be overcome to measure the IMF \cite{2013pss5.book..115K,2018PASA...35...39H}.

Deriving the IMF from theory is extremely challenging due to the numerous and complex factors that influence it \cite{annurev:/content/journals/10.1146/annurev-astro-052622-031748}, such as the thermal structure of the cold interstellar medium \cite[e.g.][]{tielens2005physics,annurev:/content/journals/10.1146/annurev-astro-071221-053453}; Supersonic turbulence \cite[e.g.][]{Hennebelle2012-jw}; Gravity, jeans instability, and gravoturbulence \cite[e.g.][]{10.1093/mnras/stu1915}; Ideal and non-ideal magnetohydrodynamics \cite[e.g.][]{Zhao2020-xp}; Tidal forces and tidal radius \cite[e.g.][]{10.1093/mnras/staa075}; Accretion \cite[e.g.][]{10.1093/mnras/stac526}. Current work relies heavily on complex numerical simulations and there is a lack of a unified framework that incorporates all these factors. This study, based on a probabilistic model, theoretically proves that the density-radius relationship of molecular clouds determines the slope. In the derivation process, some linear assumptions were used and the physical meanings represented by these assumptions were analyzed. In the future, it will be necessary to gain a deeper understanding of the aforementioned physical processes to determine the degree of deviation between the actual situation and the linear assumptions.

\section{Derivation}

Consider a molecular cloud with volume $V$. If this molecular cloud forms only one star and the mass of this star is $M$, then there are $M(V)$ and $V(M)$, with the independent variable omitted when not necessary. The expected number of stars formed by this molecular cloud is $E(V)$, thus we have

\begin{equation}\label{EV}
    E(V)=\sum_{i=1}^{\infty} i\cdot P_i(V)
\end{equation}

Here $P_i(V)$ refers to the probability of forming exactly $i$ stars within the volume. Let $\tilde{p}(V)$ be the probability of producing one star within the volume. Therefore, we have

\begin{equation}\label{pi}
    P_i(V)=\tilde{p}(V)^i \cdot (1-\tilde{p}(V))
\end{equation}

Here, it is assumed that the stellar embryos formed earlier in the molecular cloud do not affect the formation probability of the stellar embryos formed later, i.e., $\tilde{p}(V)$ remains constant. Or we can say that, assuming that all stellar embryos form simultaneously. This assumption may seem overly idealistic, but as long as the later-formed stars are metallically polluted by the earlier-formed stars, they can be distinguished observationally. Considering this, the assumption is quite acceptable. $0<\tilde{p}(V)<1$, so 

\begin{equation}
    E(V)=\sum_{i=1}^{\infty} i\cdot \tilde{p}(V)^i \cdot (1-\tilde{p}(V))=\frac{\tilde{p}(V)}{1-\tilde{p}(V)}
\end{equation}

Then

\begin{equation}\label{dp}
    \tilde{p}(V)=\frac{E(V)}{E(V)+1}
\end{equation}

Then

\begin{equation}
    P_i(V)=\frac{E(V)^i}{(E(V)+1)^{i+1}}
\end{equation}

So, the IMF is

\begin{equation}
    P(M)=\sum_{i=1} i \cdot P_i(f_1(i,V) \cdot V)
\end{equation}

This equation indicates that the probability of producing a star with mass M requires considering scenarios where a single molecular cloud directly forms one star, a molecular cloud forms two stars each with mass M, and so on. Since if a star-forming region with volume $nV$ produces n stars, their expected total mass is not necessarily M, so a correction using $f_1(i,V)$ is needed. However, here we assume $f_1(i,V)=i$, which is equivalent to assuming that when a molecular cloud with n times the volume only forms n stars, the expected mass of these n stars is equal to the mass of a star formed by a molecular cloud with one times the volume that forms only one star. Then we have

\begin{equation}\label{pm}
    P(M)=\sum_{i=1} i \cdot P_i(i V)=\sum_{i=1} i \cdot \frac{E(iV)^i}{(E(iV)+1)^{i+1}}
\end{equation}

Assume that the expected number of stars formed in a molecular cloud of volume $iV$ is much greater than 1. (Here are two scenarios in which the assumptions may not hold. First, for Population III stars, it is possible that a massive molecular cloud directly collapses into a single star. Second, in situations where gas is extremely scarce, it may happen that a gas clump forms only one star, and the mass of such a star is generally very small.) Then we have $(E(iV)+1) \sim E(iV)$, so

\begin{equation}
    P(M)=\sum_{i=1} \frac{i}{E(iV)}
\end{equation}

Assuming that when the volume of the molecular cloud increases by a factor of $i$, the expected number of stars formed also increases by a factor of $i$, hence we have $E(iV)=iE(V)$. So,

\begin{equation}
    P(M)=\sum_{i=1} \frac{1}{E(V)}=C/E(V)
\end{equation}

$i$ cannot be added to infinity, so the summation becomes a coefficient. For convenience, all constant coefficients in this article are not distinguished. $P(M)$ is the IMF, so

\begin{equation}
    \frac{C}{E(V)}=M^{-\alpha}
\end{equation}

Because $E(iV) = iE(v)$, letting $E(V) = CV$, we have

\begin{equation}
    \frac{C}{V}=M^{-\alpha}
\end{equation}

\begin{equation}
    M(V) = CV ^{1/\alpha}
\end{equation}

Under spherical symmetry, $V = C r^3$

\begin{equation}\label{a1}
    M(r) = Cr ^{3/\alpha}
\end{equation}

Note that this substitution implicitly assumes that when a molecular cloud forms only one star, the mass of the star is proportional to the volume of the molecular cloud; otherwise, it would introduce additional powers of $r$. Now consider

\begin{equation}\label{a2}
    M(r) = \int_0^r C r'^2 \rho(r') \, dr'
\end{equation}

Note that this step is equivalent to assuming a direct proportionality between the mass of the molecular cloud and the mass of the star formed when only one star is formed from the molecular cloud; otherwise, additional powers of $r$ would be introduced. Combined with equation \ref{a1} and equation \ref{a2}, we have

\begin{equation}
\frac{dM}{dr} = C r^2 \rho(r)=C' r^{3/\alpha - 1}
\end{equation}

Therefore, $\alpha$ depends on the exponent $n$ of $r$ in $\rho(r)$, with $\alpha=3/(3+n)$.

\section{Discussion and conclusion}

The result suggests that as $\alpha$ approaches 1 ($\Gamma_{\mathrm {IMF}}$ approaches 0), the density of the molecular cloud becomes uniform, which should be the minimum because it is unlikely that a molecular cloud is sparse in the center and dense at the edge. When $\alpha$ approaches 3 ($\Gamma_{\mathrm {IMF}}$ approaches -2), it indicates that the molecular cloud more closely resembles a singular isothermal sphere, which should be the upper limit. 

If we consider the scenario where E(V) is greater than but close to 1, meaning that molecular clouds are likely to collapse and form a single star, this could occur during the formation of Population III stars or in situations where there is a severe shortage of gas, such that a single molecular cloud can only form one small-mass star. In this case, the situations where $i>1$ in equation \ref{pm} can be ignored.

\begin{equation}\label{pm}
    P(M)=\frac{E(V)}{(E(V)+1)^{2}}
\end{equation}

Taking $E(V)=1+\epsilon(V)$, we have

\begin{equation}\label{pm}
    P(M)=\frac{\epsilon+1}{(\epsilon+2)^{2}}=\frac{\epsilon+1}{4\epsilon+4}=CM^{-\alpha}
\end{equation}

$\epsilon^2$ is ignored. Now $\alpha=-1$ ($\Gamma_\mathrm{IMF}=2$).

\begin{figure}[h]
    \centering
    \includegraphics[width=0.95\linewidth]{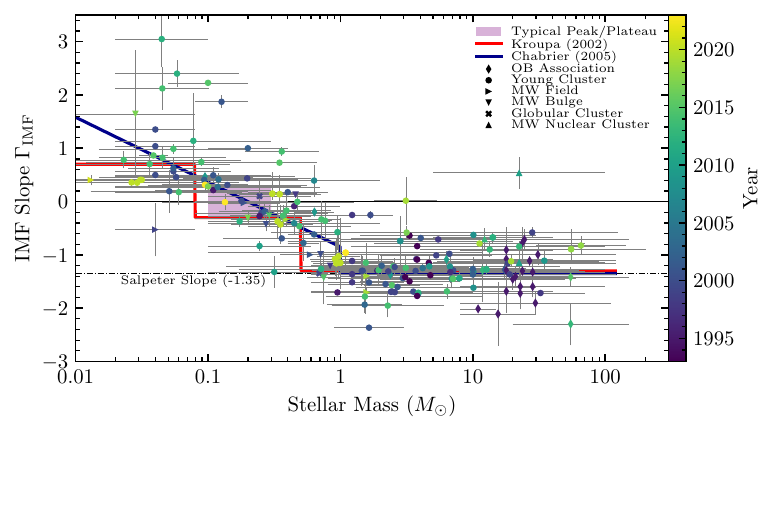}
    \caption{Created by Hennebelle \& Grudić \cite{annurev:/content/journals/10.1146/annurev-astro-052622-031748}, see https://github.com/mikegrudic/alphaplot.}
    \label{fig}
\end{figure}

Hennebelle \& Grudić \cite{annurev:/content/journals/10.1146/annurev-astro-052622-031748} summarized a large number of observed slopes and plotted Figure \ref{fig}. It can be seen that when the stellar mass is greater than 1 solar mass, $\Gamma_\mathrm{IMF}$ lies between 0 and -2, corresponding to $\alpha$ between 1 and 3; when the mass is smaller, $\Gamma_\mathrm{IMF}$ gradually approaches 2. If these very low-mass stars form in an environment where the gas is so severely deficient that E(V) approaches 1, it can explain the observed slope. As for the slope-mass relation in the figure, it may represent a transition from when E(v) is much greater than 1 to when E(V) is close to 1.

Moreover, the M in our derivation can be understood as the core mass (corresponding to the core mass function, CMF), or it can also be interpreted as the IMF, since both are within the same probabilistic framework. However, this does not imply that the theoretical predictions for their slopes are the same because when it comes to equation \ref{a1} and equation \ref{a2}, the corrections that need to be introduced if the linear assumption is not adopted are different for the CMF and the IMF. The relationship between the IMF and the CMF requires further investigation \cite{10.1093/mnras/stab844}, and the specific differences may lie in the corrections involving equation \ref{a1} and equation \ref{a2}.

\begin{acknowledgments}

This work is supported by funds from the author's institution.

\end{acknowledgments}

\bibliography{apssamp}

\end{document}